\documentstyle[aps,pre,epsfig,multicol]{revtex}
\begin{document}
\draft
\title{A Particle Model of Rolling Grain Ripples Under Waves}

\author{Ken Haste Andersen$^{1,2,3}$} \address{$^1$ Institute of
  Hydraulic Research and Water Resources, the Technical University of
  Denmark, 2800 Lyngby, Denmark. e-mail
  ken@isva.dtu.dk, www.isva.dtu.dk/$\tilde{\ }$ken,\\
  $^2$ Center for Chaos and Turbulence Studies, the Niels Bohr
  Institute, University of Copenhagen,
  2100 {\O} Copenhagen, Denmark, \\
  $^3$ Dipartimento di Fisica, Università degli Studi di Roma La
  Sapienza, Piazzale Aldo Moro 2, I-00185 Roma, Italy, tel: +39 06
  49913459, fax: +39 06 4463158.}

\date{\today} 
\maketitle
\begin{abstract}
A simple model is presented for the formation of {\em rolling grain
  ripples} on a flat sand bed by the oscillatory flow generated by a
surface wave. An equation of motion is derived for the individual
ripples, seen as ``particles'', on the otherwise flat bed. The model
account for the initial apperance of the ripples, the subsequent
coarsening of the ripples and the final equilibrium state.  The
model is related to physical parameters of the problem, and an
analytical approximation for the equilibrium spacing of the ripples
is developed.  It is found that the spacing between the ripples
scale with the square-root of the non-dimensional shear stress (the
Shields parameter) on a flat bed. The results of the model are
compared with measurements, and reasonable agreement between the
model and the measurements is demonstrated.
\end{abstract}


\begin{multicols}{2}
  In the coastal zone where the water is relatively shallow, an
  ubiquitous phenomenon is the formation of ripples in the sand. These
  ripples have been described in the seminal work of Bagnold
  \cite{bagn:46}, who called them {\em vortex ripples}. The vortex
  ripples have been studied recently from a pattern forming point of
  view \cite{steg:99,sche:99,ande:00,elle:00}. Bagnold also described
  another kind of ripples which were created only from an initially
  flat bed, as a transient phenomena before the creation of vortex
  ripples. These ripples were created by the rolling back and forth of
  individual grains on the flat bed, and were thus called {\em rolling
    grain ripples}. It is this latter class of ripples which is the
  topic of this article.
  
  The most well-known type of ripple created by oscillatory motion is
  the vortex ripple, called so because of the strong vortices created
  by the sharp crest of the ripples. These are the ripples one
  encounter when swimming along a sandy beach. The shape of the
  ripples is approximately triangular, with sides being at the angle
  of repose of the sand. The length of the ripples scale with the
  amplitude of the oscillatory motion of the water near bed, $a$. The
  dynamics of the vortex ripples and the creation of a stable
  equilibrium pattern has been recently described by \cite{ande:00}.
  
\begin{figure}[tbp]
  \begin{center}
    \epsfig{file=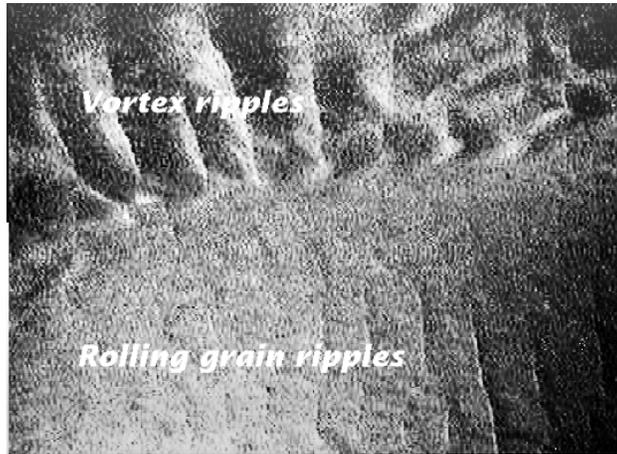,width=6cm,angle=-90}
    \caption{An example of a flat bed viewed from above, 
      with rolling grain ripples (bottom) and vortex ripples invading
      from the top. The experimental setup is a wave tank 60 cm wide
      and approximately 50 cm deep, containing sand with a median
      diameter of $0.2$ mm.}
    \label{fig:exam}
  \end{center}
\end{figure}

\begin{figure}[htbp]
  \begin{center}
    \epsfig{file=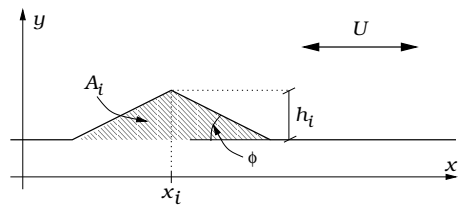,width=6cm}
    \caption{A sketch of a ripple on the flat bed with related quantities.}
    \label{fig:sketch}
  \end{center}
\end{figure}

  In contrast to the vortex ripples, the rolling grain ripples are
  rarely (if ever) found in the field, and have only been studied in
  controlled laboratory experiments, where it is possible to have a
  completely flat bed as initial conditions. In this article the term
  ``rolling grain ripple'' refer to the ridges with triangular cross
  section occurring on an otherwise flat bed (Fig.~\ref{fig:exam} and
  Fig.~\ref{fig:sketch}) which is in agreement with Bagnolds
  definition of the rolling grain ripples (see also Plate 3 in
  \cite{bagn:46}). The length of the triangular cross section is small
  (typically smaller than 1 cm), and much smaller than the spacing
  between the ridges, which again is typically smaller than the
  spacing between the vortex ripples. As the crest of the rolling
  grain ripples is rather sharp, a separation zone can be induced in
  the lee side of the ripple (a separation zone is an area where the
  flow is in the opposite direction of the main flow). This has also
  been found from numerical simulations of the flow over triangular
  ridges \cite{ande:99}.
  
  Bagnold assumed that there were no separation zones behind the
  rolling grain ripples, and this has lead to the adoption of the term
  rolling grain ripples for ripples without separation. It seems as if
  another type of ripple can emerge from a flat bed. These ripples
  initially have a sinusoidal shape, and thus no flat bed between
  them.  The ripples grow and become sharper until they are steep
  enough to induce separation, whereafter they grow to become vortex
  ripples.  From the authors own experience \cite{ande:99} and from
  \cite{foti:pers}, it seems as if which if these two kinds of ripples
  appear from a flat bed depend on the initial preparation of the bed.
  If the grains are small and the bed well packed, the rolling grain
  ripples as described by Bagnold appear (``Bagnold type rolling grain
  ripples''). If, however, the grains are large or the grains are
  carefully prepared, so as to not be packed, i.e., by letting the
  grains settle trough the water, a mode with sinusoidal ripples can
  be observed \cite{foti:pers}. These ``sinusoidal rolling grain
  ripples'' have been described by a hydrodynamic instability of the
  wave boundary layer.  The idea was originally formulated by Sleath
  \cite{slea:76}, but was developed in great detail in a series of
  papers by Blondeaux, Foti and Vittori
  \cite{blon:90,vitt:90,vitt:91,foti:95,foti:95b}.  This kind of
  ripples is not described by the model which will be developed here.

The model to be presented is a ``particle model'' in the sense that it
view each ripple as a ``particle'' which interact with its
neighbouring particles. A heuristic equation of motion is written for
each particle, taking into account the movement of the particle and
the presence of the neighbouring particles. When two particles collide
they merge and form a new, larger particle. This continues until a
steady state is achieved.

After a more detailed description of the rolling grain in section
\ref{sec:RGripples}, the heuristic model is developed in section
\ref{sec:model}. It turns out, fortunately, that the parameters
entering this model can be connected to measurable quantities (section
\ref{sec:theta}). In section \ref{sec:solution} the model is solved
numerically and analytically, and comparison with experimental
measurements of rolling grain ripples are done in section
\ref{sec:compare}. Conclusions can be found in section \ref{sec:conc}.

\section{The rolling grain ripples}\label{sec:RGripples}
An example of a flat bed with rolling grain ripples coexisting with
vortex ripples is seen in Fig.~\ref{fig:exam}. In the middle of the
picture the ripples have not yet formed and the bed is still flat,
while on the top vortex ripples are seen to invade into the flat bed.
The vortex ripples are typically nucleated from the boundaries or from
a pertubation in the bed. In the lower part of the picture the rolling
grains ripples have formed on the flat bed, and are seen as the small
bands of loose grains on top of the flat bed.

The flow over the bed created by the surface wave is oscillating back
and forth in a harmonic fashion.  This flow creates a shear stress on
the bed $\tau(t)$ which in non-dimensional form reads:
\begin{equation}
  \theta(t) = \frac{\tau(t)}{\rho (s-1) g d}~,
\end{equation}
where $\rho$ is the density of water, $s$ is the relative density of
the sand (for quartz sand in water $s = 2.65$), $g$ is the gravity and
$d$ is the mean diameter of the grains. $\theta$ is usually called the
Shields parameter \cite{shields}. When the shear stress exceeds a
critical value $\theta_c$, the grains start to move. For a turbulent
boundary layer the value of the critical Shields parameter is
$\theta_c \approx 0.06$ \cite{fred:92}. The grains which have become
loosened from the bed, start to move back and forth on the flat bed,
and after a while the grains come to rest in parallel bands. In the
lee side of each band the bed is shielded from the full force of the
flow, creating a ``shadow zone'' where the grains move slower than in
the upstream side of the bands. 

Due to this shadow zone, more grains end up in the bands than leave
the bands, and they grow until they form small ridges, the rolling
grain ripples. When the rolling grain ripples are fully developed, no
grains will be pulled loose from the bed in the space between them,
and they are stable. However, in reality the rolling grain ripples are
dominated by invading vortex ripples, which is the main reason why
they are rarely observed in Nature.

\section{A simple model}\label{sec:model}
The above scenario can be formulated mathematically by writing an
equation of motion for each grain/particle. In the following an
equation of motion for the particles are developed.  In the beginning
the particles represents the grains, but as the single grains quickly
merge, the particles most of the time represents a ripple. First the
velocity of each particle is found, assuming that the particle is
alone of the flat bed, and then the influence of the shadow zones from
neighboring particles are taken into account.

Consider $N$ particles rolling on top of a rough, solid
surface. Each particle is characterized by its position $x_i$ and its
height $h_i$ (see Fig.~\ref{fig:sketch}). As the ripples are
triangular, the area of each particle $A_i$ and their heights are
related as:
\begin{equation}
  h_i = \sqrt{A_i \tan \phi}~,
  \label{eq:height}
\end{equation}
where $\phi$ is the angle of repose of the sand (33$^\circ$).

\begin{figure}
  \begin{center}
    \epsfig{file=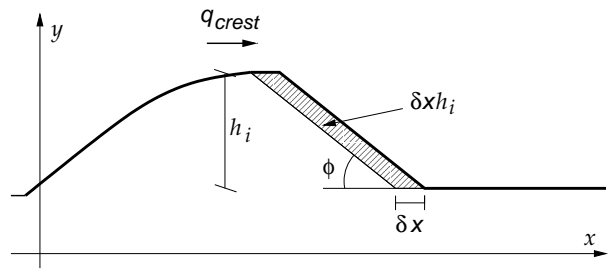, width=8cm}
    \caption{Geometrical illustration of the quantities involved
      in the derivation of the one-over-h law.}
    \label{fig:one-over-h}
  \end{center}
\end{figure}

The ripple moves back and forth slower than a single grain, according
to the ``1/height'' law. This law is well known in the study of dunes
in the desert \cite{nish:98} or sub-aqueous dunes \cite{fred:96}, and
can be illustrated by a simple geometrical argument. Suppose there is
a flux of sand over the crest of a ripple or a dune $q_{crest}$
(Fig.~\ref{fig:one-over-h}).  To make the ripple move a distance
$\delta x$ an amount of sand $h \delta x$ is needed.  As the sediment
flux is amount of sand per unit time, the velocity of the ripple is
$u_{ripple}\!=\!q_{crest} / h \propto 1/h$. If the height of the
initial particles (the single grains) is assumed to be equal to the
grain diameter $d$, the velocity of the particles can be related to
that of the single grains as:
\begin{equation}
  u_i = \frac{d}{h_i} U_g \sin(\omega t)~,
\end{equation}
where $U_g$ is the velocity amplitude of the motion of a single grain
and $\omega$ is the angular frequency of the oscillatory motion.

\begin{figure}[tb]
  \begin{center}
    \epsfig{file=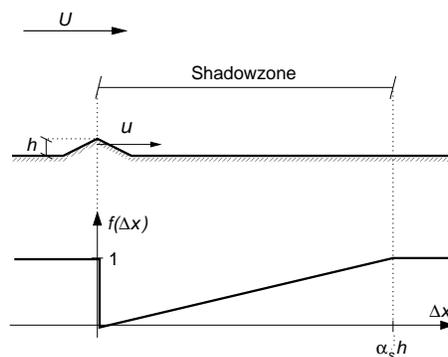, width=6cm}
    \caption{An illustration of the shadow zone of one ripple, in the
      part of the wave period where the flow is from the left to the
      right. Below is seen the function $f(\Delta x)$ used to describe
      how a particle is slowed down when it enters the shadow zone of
      another particle.}
      \label{fig:shadowzone}
  \end{center}
\end{figure}

The flow moves back and forth and makes the particles move accordingly
on top of the bed. In the wake of each particle/ripple there is a
shadow zone (Fig.~\ref{fig:shadowzone}), which is the area behind the
particle where the absolute value of the shear stress is smaller than
it would be on a flat bed. The length of the shadow zone is therefore
larger than the length of the separation bubble formed by the particle
(note that the shadow zone would be present even in the absence of
separation). If the shadow zone is much smaller than the amplitude of
water motion, the flow in the lee side of the ripple can be assumed to
have sufficient time to become fully developed. The fully developed
flow over a triangle is similar to that past a backward facing step in
steady flow, which has been extensively studied (see e.g. Tjerry
\cite{tjer:95}). In that case the relevant quantities, i.e., the
length of the separation bubble, the length of the wake etc., scale
with the height of the step. As a first assumption, the shadow zone is
therefore assumed to have a length which is proportional to the height
of the particle: $\alpha_s h_i$. If a particle enters the shadow zone
of another particle, it is slowed down according to the distance
between the particles. This means that the actual velocity of a
particle is $u_i f(\Delta x)$ where $\Delta x$ is the distance between
the grain and the nearest neighbor upstream, $u_i$ is the velocity of
the particle outside the shadow zone and $f$ is a function determining
the nature of the slowing of the motion of the particle.  A simple
linear function is used, as shown in Fig.~\ref{fig:shadowzone}. The
exact form of the function $f$ is not crucial, as will be evident
later, the important parameter being the extent of the shadow zone as
determined by $\alpha_s$.

It is now possible to write the equations of motion for the particles
as a system of coupled ODEs:
\begin{equation}
  \dot{x_i} = \frac{d}{h_i} u_g(t) %
  \underbrace{f\!\left(\frac{x_i - x_{i-1}}{\alpha_s
        h_{i-1}} \frac{u(t)}{|u(t)|}\right)}_{\mathrm{positive\ half\
      period}}%
  \underbrace{f\!\left(\frac{x_i - x_{i+1}}{\alpha_s
        h_{i+1}} \frac{u(t)}{|u(t)|}\right)}_{\mathrm{negative\ half\
      period}}
  \label{eq:model}
\end{equation}
for $i=1,\ldots,N$, where $u_g(t) = U_g \sin(\omega t)$. The motion of
a particle is thus made up of three parts {\em (i)} the motion of the
single undisturbed particle, {\em (ii)} the effect of the shadow from
the particle to the left $(i-1)$, which might affect particle $i$ in
the first half period and {\em (iii)} the effect of the shadow of the
particle to the right $(i+1)$ in the second half period.

When lengths are scaled by the diameter of the grains and time by the
frequency $\omega$, it is possible to identify the three relevant
dimensionless parameters of the model:
\begin{description}
\item{$\alpha_s$,} the length of the shadow zone divided by the height.
\item{$a_g/d$,} the amplitude of the motion of a single grain, divided
  by the grain diameter ($a_g = \omega U_g$).
\item{$\lambda_i/d$,} the initial distance between the grains;
  $\lambda_i = L/N$, where $L$ is the length of the domain and $N$ is
  the initial number of grains.
\end{description}

\subsection{Relation to physical quantities}\label{sec:theta}
Even though the model seem quite heuristic, the parameters entering the
model, $a_g/d$, $\lambda_i/d$ and $\alpha_s$ can be related to
physical parameters describing the flow and the properties of the
grains. The line of arguments presented here closely follows those
used to derive the flux of sand on a flat bed (the bed load), as can
be found in text books, i.e., \cite{fred:92}, or in \cite{kova:94}.
First the velocity of a single grain will be derived, from which
$a_g/d$ can be inferred. Thereafter the initial number of grains in
motion is found, from which follows $\lambda_i/d$. Finally the length
of the shadow zone $\alpha_s$ is discussed.

\subsubsection{The velocity of the grains}
The velocity of the grain can be found by considering the force
balance on a single grain lying on the flat bed. The grain is subject
to a drag force proportional to the square of the relative flow
velocity $u_r = u_{nb} - u_g$ where $u_{nb}$ is the velocity near the
bed and $u_g$ is the velocity of the grain:
\begin{equation}
  F_d = \frac{1}{2} C_D \rho A |u_r| u_r,
\end{equation}
where $A$ is the area of the grain and $C_D$ is a drag coefficient.
The numerical signs is used to obtain the right sign of the force. The
velocity profile in the vicinity of the bed is supposed to be
logarithmic. As shown in \cite{oldmeas} this is a reasonable
assumption except when the flow reverses. However, during reversal the
velocities are anyway small and the accuracy is of minor importance.
The logarithmic profile over a rough bed can be written as (eg.
\cite{fred:92}):
\begin{equation}
  u(y) = \frac{u_f}{\kappa} \ln\left(\frac{30 y}{k_N} \right),
    \label{equ:wall}
\end{equation}
where $\kappa = 0.41$ is the von K{\'a}rm{\'a}n constant, $u_f \equiv
\sqrt{\tau/\rho}$ is the friction velocity and $k_N$ is the Nikuradse
roughness length. It is then possible to find the near bed velocity as
the velocity at $y\!\approx\!1/2 d$:
\begin{equation}
  u_{nb} = \xi u_f 
\end{equation}
where the constant $\xi$ can be determined from
Eq.~(\ref{equ:wall}) by assuming $k_N = d$. Opposing the drag on the
grain is the friction of the bed:
\begin{equation}
  F_f = -\mu W
\end{equation}
where $\mu$ is a friction coefficient and $W = \rho g (s-1)d^3\pi/6$
is the immersed weight of the grain. By making a balance of forces,
$F_d + F_f =0$, the velocity of the grain can be found:
\begin{equation}
  u_g = \xi u_f \left( 1 - \sqrt{\left| \frac{\theta_c}{\theta}
      \right| } \right),
\end{equation}
where
\begin{equation}
  \theta_c = \frac{4 \mu}{3 C_D \xi^2}~,
\end{equation}
is the critical Shields parameter. Usually $\mu = \tan \phi \approx
0.65$ \cite{fred:92}.

\subsubsection{The initial spacing of grains}
Now that the velocity of the grains has been calculated, there still
remains to determine the number of grains per area $n$ in motion. To
this end a small volume of moving sand at the top the flat bed is
considered.  The balance of the forces acting on this volume is
written as:
\begin{equation}
  \tau_b = \tau_G + \tau_c.
  \label{equ:bal}
\end{equation}
The interpretation of the terms is as follows: The parameter $\tau_b$
is the shear stress on the top of the bed load layer.  It is assumed
that this is equal to the shear stress on a fixed flat bed.  $\tau_G$
is the stress arising from the inter-granular collisions, giving rise
to ``grain-stresses'' \cite{kova:94}, modeled as: $\tau_G = n \mu W
$.  It is assumed that the inter-granular stress absorbs all the
stress, except the critical stress $\tau_c$; this is the so-called
``Bagnold hypothesis'' \cite{kova:94}.  Making Eq.~(\ref{equ:bal})
non-dimensional by dividing with $\rho (s-1)gd$, the number of grains
in motion is found as:
\begin{equation}
  n = \frac{6}{\pi d^2 \mu}(\theta - \theta_c),
  \label{equ:n}
\end{equation}
If $\theta < \theta_c$, then no grains are in motion and $n = 0$. $n$
can also be viewed as the initial density of grains, and by assuming a
square packing of the grains the initial distance between the grains
becomes:
\begin{eqnarray}
  \frac{\lambda_i}{d} &=& \frac{1}{\sqrt{n} d}\\
                      &=& \sqrt{\frac{\pi \mu}{6 (\theta -
                          \theta_c)}}.
  \label{equ:lambda_f}
\end{eqnarray}

\subsubsection{The length of the shadow zone}
\label{sec:length}
\begin{figure}[htbp]
  \begin{center}
    \epsfig{file=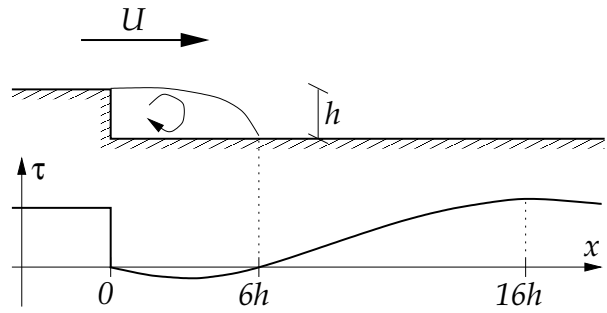, width=8cm}
    \caption{A sketch of the shear stress on the bed during a steady
      flow over a backward facing step.} 
    \label{fig:backward}
  \end{center}
\end{figure}

The last parameter $\alpha_s$, which characterize the length of the
shadow zone, is a bit harder to estimate accurately. By exploring the
analogy with the backward facing step which was suggested in section
\ref{sec:model}, we can get some bounds for $\alpha_s$. In the
backward facing step there is a zone with flow separation which
extends approximately 6 step heights from the step, see
Fig.~\ref{fig:backward}. After approximately 16 step heights there is
a point where the shear stress has a small maximum. It follow that the
length of the shadow zone should be longer than the separation zone,
but shorter that the point of the maximum in the shear stress, i.e.,
$6 < \alpha_s < 16$.
 

\subsection{Numerical and analytical solutions of the model}
\label{sec:solution}
In the following section the behavior of the model is examined. To
study the detailed behavior the set of coupled ordinary differential
equations (\ref{eq:model}) is integrated numerically. It is seen that
the model reaches a steady state and an analytical expression for the
spacing between the ripples in the steady state is developed.
  
The numerical simulations in this section are based on a simple
example with $a_g/d\!=\!35$ and $\lambda_i/d\!=\!3.23$ and $\alpha_s$
is set to 10. As initial condition all particles have an area of $1.0\ 
\pm$ 10 \%, to add some perturbation. The initial number of particles $N$
in this example is 800. 

\begin{figure}[tb]
  \begin{center}
    \epsfig{file=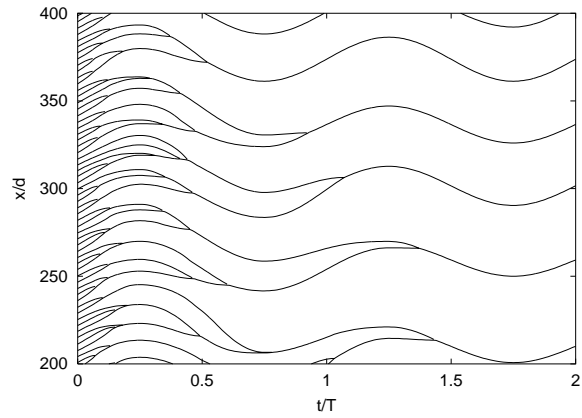, width=8cm}
    \caption{Zoom of the movement of the particle in the first two wave
      periods.  $a_g/d = 35$, $\lambda_f = 3.23$ and $\alpha_s =
      10.0$.}
    \label{fig:RG1}
  \end{center}
\end{figure}

\begin{figure}[tb]
  \begin{center}
    \epsfig{file=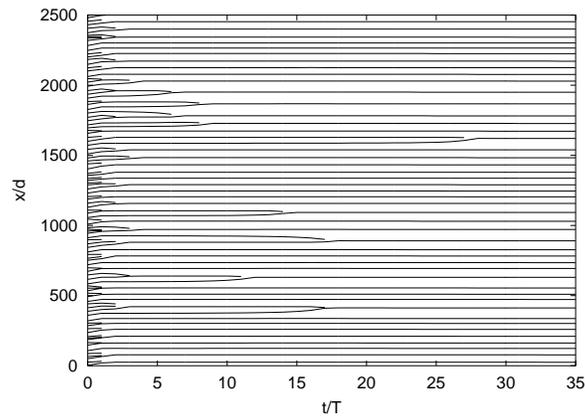, width=8cm}
    \caption{The development of the particles until steady state is
      reached. The lines show the positions of the particles at the
      end of each period.}
    \label{fig:RG2} 
  \end{center}
\end{figure}

In the first few periods a lot of grains is colliding and merging
(Fig.~\ref{fig:RG1}). As the ripples are formed and grow bigger the
evolution slows down, until a steady state is reached
(Fig.~\ref{fig:RG2}). The spacing between the ripples in the steady
state show some scatter around the average value, which is also
observed in experiments. The variation in the average spacing of the
steady state $\lambda_{eq}$ for realizations with different initial
random seed, turned out to be on the order of $1/N_{eq}$ where
$N_{eq}$ is the final number of ripples.

To study the behavior of the average spacing of the ripples
$\lambda_{eq}$, a number of simulations were made where the parameters
were varied one at a time.  Each run was started from the initial
disordered state.

\begin{figure}[tb]
  \begin{center}
    \epsfig{file=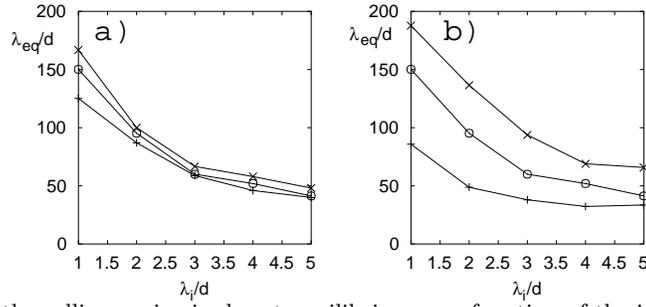, width=8.5cm}
    \caption{The spacing between the rolling grain ripples at 
      equilibrium as a function of the initial density of grains. The
      basic example is shown with the circles: $a_g/d = 35$ and
      $\alpha_s = 10.0$.  In a) the value of $a_g/d$ is varied, for
      pluses $a_g/d = 20$, crosses $a_g/d = 50$. In b) $\alpha_s$ is
      varied; pluses: $\alpha_s = 7$, crosses: $\alpha_s = 13$.}
      \label{fig:RGlength}
  \end{center}
\end{figure}

Changing $a_g/d$ only results in a minor change in the spacing of the
ripples (Fig.~\ref{fig:RGlength}a). The final spacing between the
ripples does depend on the length of the shadow zone $\alpha_s$; the
longer the shadow zone, the longer the ripples
(Fig.~\ref{fig:RGlength}b).  This can be used to estimate the average
equilibrium spacing between the ripples.  When the distance between
two ripples is longer than the shadow zone of the ripples, they are no
longer able to interact. This gives:
\begin{equation}
  \lambda_{eq} >  \alpha_s h_{eq},
  \label{eq:lower}
\end{equation}
where subscript $eq$ denotes average value at equilibrium. However if
two ripples have a spacing just barely shorter than Eq.
(\ref{eq:lower}), they will be able to interact and eventually they
will merge. One can therefore expect to find spacings up to
$\lambda_{eq}\!<\!2\alpha_s h_{eq}$. Assuming that the average length
is in between the two bounds one get:
\begin{equation}
  \lambda_{eq} = \gamma \alpha_s h_{eq},\quad 1<\gamma<2~.
  \label{eq:aa}
\end{equation}
where $\gamma$ can be found by comparing the results from the full
simulations with Eq.~(\ref{eq:aa}). The height of the ripples at
equilibrium can be found by splitting the initial number of particles
evenly onto the equilibrium ripples. Then the average area at
equilibrium is $A_{eq}\!=\!\lambda_{eq}/\lambda_i d^2$ and from
Eq.~(\ref{eq:height}) follows that the height is
$h_{eq}/d\!=\!\sqrt{\tan\phi\,\lambda_{eq}/\lambda_i}$, which gives an
average equilibrium spacing:
\begin{eqnarray}
  \frac{\lambda_{eq}}{d} &=& 
  \gamma^2\frac{  \alpha_s^2 d\tan\phi}{\lambda_i}
  \nonumber \\
   &=& \alpha_s^2 \gamma^2\sqrt{\frac{6\tan\phi}{\pi}} \sqrt{\theta - \theta_c}~.
   \label{equ:equilibrium}
\end{eqnarray}
The equilibrium spacing is therefore found to be proportional to
$\sqrt{\theta\!-\!\theta_c}$ with the constant of proportionality
being made up of $\alpha_s$, $\gamma$ and various geometrical factors.
All the quantities related to the dynamical evolution of the ripples,
i.e., the velocity of the ripples, the shape of the function $f(\Delta
x)$ etc.~does not enter into the expression.
 
\subsection{Comparison with experiments}\label{sec:compare}
The only parameter that has not been accurately determined is
$\alpha_s$. The value of this parameter can be estimated by comparison
with measurements.

In 1976, Sleath made a series of experiments, measuring spacing
between rolling grain ripples \cite{slea:76}. The ripples were formed
on a flat tray oscillating in still water using sand of two different
grain sizes: 0.4 mm and 1.14 mm. To compare with the experiments the
value of $\theta$ need to be calculated. $\theta$ reflect the number
of grains in motion, and it seems natural to use the maximum value
during the wave period, $\theta_{max}$. To estimate $\theta_{max}$ the
shear stress on the bed has to be estimated. The maximum shear stress
on the bed, $\tau_{max}$ during a wave period can be found using the
concept of a constant friction factor $f_w$ (\cite{jons:76}):
\begin{equation}
  \tau_{max} = \frac{1}{2} \rho f_w u_{max}^2,
\end{equation}
with $u_{max}$ being the maximum near-bed velocity. The friction
factor can be estimated using the empirical relation \cite{fred:92}:
\begin{equation}
  f_w = 0.04 \left( \frac{a}{k_N} \right)^{-0.25}
\end{equation}
where $a = u_{max}\omega$ and $k_N \simeq d$.

The range of Shields parameters in the experiments was found in this
way to range from the critical Shields parameter to $\theta = 0.42$.
For the high Shields parameters the rolling grain ripples were
reported to be very unstable, and to quickly develop into vortex
ripples. In these cases, the measured ripple spacing then reflects the
spacing between the rolling grain ripples before they developed into
vortex ripples \cite{slea:76}.

\begin{figure}[tb]
  \begin{center}
    \epsfig{file=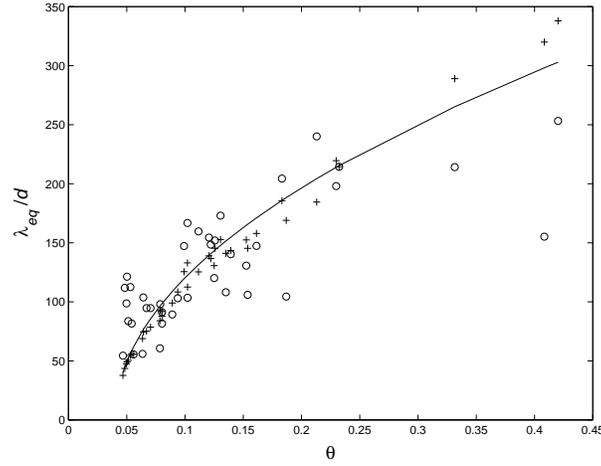, width=8cm}
    \caption{Comparison between the measured wave lengths of rolling
      grain ripples (circles), with results from the model (pluses),
      and from Eq.~(\protect\ref{eq:aa}) with $\gamma \!=\!1.40$ (the
      line). The value of $\alpha_s$ is $15.0$.  The critical Shields
      parameter is 0.04 and $\mu\!=\!0.65$.}
    \label{fig:Compare}
  \end{center}
\end{figure}

In Fig.~\ref{fig:Compare} the experimental results are compared with
runs of the model using $\alpha_s = 15.0$ (the reason for this
particular value will be clear shortly) and $N=10000$. By fitting all
the runs to Eq.~(\ref{eq:aa}) it was found that $\gamma=1.40$.  The
results using Eq.~(\ref{equ:equilibrium}) and $\gamma=1.40$ are shown
with a line.

First of all, it is seen that Eq.~(\ref{equ:equilibrium}) predicts the
results from the full model Eq.(\ref{eq:model}) well. The
correspondence between the model and the experiments is reasonable,
but there are some systematic discrepancies, which will be discussed.

There are some few points with small ripple spacing for which the
model does not fit the measurements.  These measurements has a Shields
parameter very near the critical (i.e., just around the onset of grain
motion), which implies some additional complications. The grains used
in the experiment were not of a uniform size; rather they were part of
a distribution of grain sizes, and the grain size reported is then the
median of the distribution, $d_{50}$.  The Shields parameter is
calculated using the median of the distribution, but actually one
could calculate a Shields parameter for different fractions of the
distribution, thus creating a $\theta_{10}$, a $\theta_{50}$ etc..
When $\theta_{50}$ is smaller than then the critical Shields
parameter, $\theta_{10}$ might still be higher than the critical
Shields parameter. This implies that grains with a diameter smaller
than $d_{50}$ will be in motion, while the larger grains will stay in
the bed. As only $d_{50}$ is used in the calculation of the
equilibrium ripple spacing, the distance between the grains
$\lambda_i$ will be overestimated near the critical Shields parameter,
where the effect of the poly-dispersity is expected to be strongest.
An overestimation of $\lambda_i$ will lead to an under-prediction of
the ripple length, which is exactly what is seen in
Fig.~\ref{fig:Compare}.

There are also three points from the experiments taken at very large
Shields parameters, which seems to be a bit outside the prediction of
the model. As already mentioned, these points are probably doubtful
because of the very fast growth of vortex ripples. It is therefore
reasonable to assume that vortex ripples invaded the rolling grain
ripples before these had time to reach their full length.

To find a reasonable value of $\alpha_s$ Eq.~(\ref{equ:equilibrium})
was fitted to the experimental points. To avoid the points which might
be of doubtful quality, as discussed above, only the points in the
range $0.075\!<\!\theta\!<\!0.3$ were used. This gave the value of
$\alpha_s\!=\!15.0$, in agreement with the qualitative arguments in
section \ref{sec:length}.
 
\section{Discussion of the results}\label{sec:discuss}
From the comparison of the model with measurements it seems as if the
model confidently reproduces the experiments.  

In the model the number of grains in motion is constant (even though
the number of particles changes). In an experimental situation,
however, new grains might be lifted from the bed and added to the
initial number of grains in motion. As the part of the flat bed
between the ripples is covered by the shadow zones of the particles,
these stretches will be shielded from the full force of the flow, and
only very slowly will new grains be loosened here. This small addition
of new grains will result in a slow growth of the rolling grain
ripples, such that eventually grow into vortex ripples. This slow
growth is very well illustrated by recent measurements \cite{steg:99}, 
but not covered by the present model.

\section{Conclusion}\label{sec:conc}
In conclusion, a model has been created which explain the creation and
the equilibrium state of rolling grain ripples of the type described
by Bagnold. The final distance between the ripples is proportional to
$\sqrt{\theta - \theta_c}$. The model has been compared with
measurements with reasonable agreement.

\acknowledgements It is a pleasure to thank Tomas Bohr, Clive
Ellegaard, Enrico Foti, J{\o}rgen Freds{\o}e and Vachtang Putkaradze
for useful discussions. I also wish to thank the anonymous referees
for constructive criticism.


\end{multicols}
\newpage

\end{document}